\documentclass[sigplan,screen]{acmart}

\usepackage{listings}
\usepackage{common}
\usepackage{color}
\usepackage{etoolbox}
\usepackage{fontspec}
\usepackage{xeCJK}
\usepackage{simplebnf}
\newcommand{\cw}[1]{\textcolor{black}{#1}}

\definecolor{keywordcolor}{rgb}{0.7, 0.1, 0.1}   
\definecolor{tacticcolor}{rgb}{0.0, 0.1, 0.6}    
\definecolor{commentcolor}{rgb}{0.4, 0.4, 0.4}   
\definecolor{symbolcolor}{rgb}{0.0, 0.1, 0.6}    
\definecolor{sortcolor}{rgb}{0.1, 0.5, 0.1}      
\definecolor{attributecolor}{rgb}{0.7, 0.1, 0.1} 

\usepackage{juliamono}

\AfterEndEnvironment{lstlisting}{\normalcolor}
\graphicspath{{./assets/}}

\AtBeginDocument{%
  }

\setcopyright{cc}
\setcctype{by}
\acmDOI{10.1145/3830435.3830954}
\acmYear{2026}
\copyrightyear{2026}
\acmISBN{979-8-4007-2867-9/2026/08}
\acmConference[FARM '26]{Proceedings of the 14th ACM SIGPLAN International Workshop on Functional Art, Music, Modelling, and Design}{August 24--29, 2026}{Indianapolis, IN, USA}
\acmBooktitle{Proceedings of the 14th ACM SIGPLAN International Workshop on Functional Art, Music, Modelling, and Design (FARM '26), August 24--29, 2026, Indianapolis, IN, USA}
\acmSubmissionID{icfpws26farmmain-p93-p}
\received{2026-06-02}
\received[accepted]{2026-06-18}

\begin{document}

\title{Prismriver: Formalization of Music Theory and Algorithmic Composition in Lean 4}

\author{Leni Aniva}
\orcid{0000-0002-6033-9140}
\affiliation{%
  \institution{Stanford University}
  \department{Computer Science Department}
  \city{Stanford}
  \country{USA}
}
\email{aniva@stanford.edu}

\author{Claire Wang}
\orcid{0009-0002-9321-0877}
\affiliation{%
  \institution{University of Pennsylvania}
  \city{Philadelphia}
  \country{USA}
}
\email{cdwang@seas.upenn.edu}

\renewcommand{\shortauthors}{L. Aniva and C. Wang}

\begin{abstract}
  Music theory obeys a rich set of mathematical rules and symmetries. These
symmetries follow mathematical structures which can be verified and expressed
in the precise language of a proof assistant. In this paper, we present
Prismriver, a formalization library of music theory in Lean 4. We use Prismriver to generalize beyond existing work that assumes equal temperament tuning. We also discuss modelling counterpoint music theory with Prismriver. By formalizing music
theory in Lean 4, we open the door to verifiable algorithmic composition and
accompaniment generation. Prismriver also has a custom DSL integrated with MusicXML
exports to interoperate with other music software. Prismriver can be used to compose music with Lean, using monadic composition primitives.

\end{abstract}



\maketitle

\section{Introduction}

Music theory studies the structures underlying composition, which can have mathematical parallels \citep{tymoczko2006geometry}. For example, when representing pitch classes as integers in $\mathbb{Z}/12\mathbb{Z}$, transposing the C major chord $\{0,4,7\}$ up by five semitones yields the F major chord $\{5,9,0\} = \{0+5,\ 4+5,\ 7+5\} \bmod 12$, showing how musical transposition corresponds to translation \citep{tymoczko2006geometry}. The rules and symmetries in music theory can be stated precisely and proven as opposed to only described informally. Mathematical music theory already consists of proving theorems on music theory structures \citep{crans_musical_2008} though such proofs are written in natural language, which can only be checked by hand and peer review. Checking handwritten proofs can lead to missed errors and lack of mechanical reuse of results. Interactive theorem provers are software systems for writing proofs that can be verified by a computer, eliminating the inefficiencies of handwritten proofs \citep{harrison_history_2014}. The Lean \citep{noauthor_lean_nodate} theorem prover is a functional language and theorem prover accompanied by Mathlib \citep{noauthor_mathematics_nodate}, a Lean library containing the largest collection of math proofs. The Lean language is highly expressive for defining custom domain specific languages given its support for rich custom abstractions, extensible syntax, and metaprogramming. When formalizing music theory, users may benefit from using existing musical primitives and the ability to use Mathlib with their proofs.

We present Prismriver, a Lean 4 library for the representation, analysis, and
algorithmic composition of music.\footnote{The code of this project can be found
at \url{https://codeberg.org/aniva/Prismriver}} Lean is both a proof assistant and a general-purpose programming language, which enables us to build a versatile library that both proves theorems about musical structure and performs I/O to compose music. Lean can also prove properties about composition algorithms such as that of first species counterpoint.

\cw{Our contributions with Prismriver are the following}:
\begin{enumerate}
\item Formalization of music theory including pitches, accidentals, scales,
	chords, durations, and parts.
\item Compatibility with xenharmonic scales including quartertones, non-equally
	tempered tuning systems, and tuning systems with a fundamental interval that
	is not the octave.
\item Provable reduction of music properties to 12-tone equal tempered scale.
\item Monadic algorithmic analysis and composition of music.
\item LilyPond-like DSL for the easy input of music and MusicXML export for
	compatibility with other music software.
\end{enumerate}

\section{Related Work}
Domain-specific languages (DSLs) for music have a long history, with applications for music synthesis, composition, and performance dating back to the 1950s \citep{loy_programming_1985}. Live coding \citep{nilson_live_2007}, an interactive way of composing music is achieved with DSLs such as Tidal \citep{tidal} and Strudel \citep{strudel}. Live coding performances are not exclusive to music performances since there are DSLs such as Hydra \citep{hydra} which synthesize performance visuals. Tidal and Strudel both provide functional primitives where users can input their specifications while automatically synthesizing patterns of music. Unlike Tidal and Strudel which provided interactive shells to play composed music, Prismriver's compositions are exported to MusicXML \citep{musicxml}, a universal file format for digital sheet music, which can then be played using Alda \citep{alda}, a music composition and playback programming language. Prismriver compositions can also be verified since the compositions are implemented in Lean. Currently Prismriver only supports music composition, and generated visualization accompanying the composition is subject to future work. 

Type systems can encode and enforce musical specifications \citep{magalhaes2011functional,szamozvancev_well-typed_2017,cong_demo_2019}. HarmTrace uses Haskell generalized algebraic datatypes (GADTs) to encode rules of tonal harmony \citep{magalhaes2011functional}. HarmTrace uses GADTs to express hierarchy since in harmony theory, not every chord in a sequence is equally important \citep{magalhaes2011functional}. HarmTrace is a tool for harmony analysis given a sequence of chord labels while Prismriver is a programmatic composition and analysis library. Mezzo is a Haskell embedded domain-spcific language for music composition, that encodes classical music theory rules using type-level constraints \citep{szamozvancev_well-typed_2017}. A Mezzo composition that violates encoded rules fails to compile and custom type errors provide feedback on the related rule, while in Prismriver, rules can be proven in Lean theorems and a violation makes a proof incomplete. MusicTools is an Agda library for analyzing and synthesizing music that uses dependent types to encode musical rules \citep{cong_demo_2019}. MusicTools is limited to the chromatic scale while Prismriver can support arbitrary tunings.

Lulu \citep{palalansouki_iehalitylulu_2026} is a Lean music composition tool which converts Lean code to MusicXML \citep{musicxml}, allowing users to write compositions programmatically in Lean. Lulu also renders MusicXML in the Lean InfoView, allowing users to inspect their Lean written compositions while using the text editor. However unlike Prismriver, Lulu does not have generalized music primitives for arbitrary tunings.  


\section{Theory Representation}

This section presents Prismriver's representation of music theory. First, we introduce the theory representation primitives in Prismriver such as scales and intervals. Then we generalize the equal temperament tuning dihedral group action of transpose and inversion from \citep{crans_musical_2008} by defining the transpose action which acts on arbitrary scales. We prove that under equal temperament tuning, the transpose action is equivalent to the $D_{12}$ dihedral group action on triads. Finally we discuss the representation of time and notes.


Before we can represent music structures, we need to introduce Lean's data
structures. A \lstinline{structure} \citep{lean_structure} is a product type of several fields, similar
to the \lstinline{struct} in Rust \citep{rust_struct} or C \citep{kernighan1988c}. \lstinline{class} is used to represent polymorphism in Lean. A \lstinline{class} is an interface which specifies a
list of functions and properties associated with a type parameter \citep{lean_typeclass}. An
instantiation of this list of functions is an \lstinline{instance} \citep{lean_instance}. For example,
\lstinline{Add} class represents a generic addition operation \lstinline{add}.
The natural number type \lstinline{Nat} \citep{lean_nat} has an instance for the \lstinline{Add}
class, for additions on natural numbers, where \lstinline{add} is a binary
function on its type parameter. Certain instances such as \lstinline{ToString} \citep{lean_toString}
could be implemented automatically using Lean macros \citep{lean_macros}. This implementation is
triggered by \lstinline{deriving} \citep{lean_deriving}. A class \lstinline{B} may be based on another
class \lstinline{A}, where instancing \lstinline{A} is a prerequisite to
instancing \lstinline{B}. In Lean, \lstinline{B} \keyword{extends} \citep{lean_extends}
\lstinline{A}.

Prismriver is designed to accommodate tuning systems and scales beyond the equal
temperament heptatonic scale in Western classical music. The most general
structure for pitch in Prismriver is a \keyword{pseudo-scale}, which contains \keyword{pitch type} $P$ with no additional restrictions.
For instance, real number frequencies  are a possible pseudo-scale
$P := \mathbb R$. In synthesizer applications, this would be represented by
floating point frequencies or MIDI pitches. A \keyword{scale} is a pitch type
$P$ with an \keyword{interval type} $I$, where the intervals $I$ form a group
action on $P$, and $I$ is an abelian additive group. i.e. $P$ is an $I$-torsor
(Listing~\ref{lst:scale-lean}). Throughout Prismriver, we use group actions
\citep{noauthor_mathlibalgebragroupactionbasic_nodate} as a  general interface
for arithmetic operations between each of our classes. For example, group
actions allow us to calculate distances between a \keyword{pitch} and an
\keyword{interval}.
\begin{lstlisting}[language=lean,float=htbp,
    caption={Scale and tuning abstractions in Lean. (\texttt{Prismriver/Repr/Scale.lean})},
    label={lst:scale-lean}]
/-- A scale is a set of pitches -/
class PseudoScale (P : Type) where
  name : String
/-- We make no distinction between scales and tuning systems. They are represented by the same class. For example, a tuning system could be represented as a scale with raw frequencies, and any abstract scale could be lifted into the raw frequency scale. This would represent tuning. -/
class Tuning (P1 P2 : Type) (src : PseudoScale P1) (dst : PseudoScale P2) where
  liftPitch : P1 → P2
/-- A scale with a repeating fundamental interval. Each pitch in the scale is represented as a tone along with a multiple of the fundamental interval. -/
class Scale (P I : Type) extends PseudoScale P, Add I, HAdd P I P, HSub P P I,
    SMul Int I, Neg I where
/-- The fundamental interval (usually an octave) -/
    fundamental : I
/-- List all notes in the 0th interval. e.g. For C major, this would be C,D,E,F,G,A,B -/
    pitches : List P
\end{lstlisting}

At this level, it makes sense to discuss the
\keyword{fundamental intervals} on a scale (e.g. the octave) and transposition.
For instance, the \keyword{12-tone equal temperament scale} is where the pitch
type is $P_\sET := \mathbb Z$ with a fundamental interval $1_\sET := 12$, since
there are 12 semitones in an octave. Following MIDI conventions, we associate
$\pitch{A_4}$ with the pitch $p_\sET := 69$. Thus
$p_{\pitch{A_4}} + 1_\sET = p_{\pitch{A_5}}$. In this scale, enharmonic equivalence
ensures $p_{\pitch{A_5\sharp}} = p_{\pitch{B_5\flat}}$.

A \keyword{tuning system} maps one pseudo-scale to another. In this case,
commonly $p_\sET \mapsto 440 \cdot 2^{p_\sET/12 - 4}$ corresponds to a pitch
standard of $\pitch{A_4}$ being $440\text{ Hz}$. We accommodate alternative
tuning schemes where the frequency of $\pitch{A_4}$ is another number.

Previous works such as \citep{crans_musical_2008} stop at 12-tone equal-tempered
tuning.  We move one step further. To represent western classical music (Listing~\ref{lst:classical-lean}), the
pitch type $P_\sCl$ is a tuple $(n, a)$, where $n \in \mathbb Z$ representing
the name of the note (e.g.  $\pitch{C_4}$), and $a \in \mathbb Z$ is the accidental
measured in semitones. We intentionally separate the pitch into two components,
since under non-equally tempered tuning systems, normally enharmonic pitches may
not be enharmonic. e.g. it is not necessarily true that
$\pitch{C_4\sharp} = \pitch{D_4\flat}$. Under this system, there are infinitely
many enharmonic pitches with different names. e.g.
$\cdots = \pitch{A_3\sharp^5} = \pitch{B_3\sharp^3} = \pitch{C\sharp} = \pitch{D\flat} = \pitch{E\flat^3} = \cdots$.
It is also conducive to the analysis of music that enharmonic tones can have
distinct names. e.g. a $\pitch{C}$-major chord is
$(\pitch{C}, \pitch{E}, \pitch{G})$ and not
$(\pitch{C}, \pitch{F\flat}, \pitch{G})$ even if they are enharmonic.
\begin{lstlisting}[language=lean,float=htbp,
    caption={Western classical Accidental, Pitch, Interval, and Tone abstractions in Lean. \\ (\texttt{Prismriver/Repr/Classical.lean})},
    label={lst:classical-lean}]
structure Accidental where semitones : Int := 0 deriving BEq, Inhabited, Ord

structure Pitch where
  name : Int
  acc : Accidental := .natural
  deriving BEq, Inhabited, Ord

structure Interval where
  name : Int
  semitones : Int
  deriving BEq, Inhabited

 structure Tone where
  name : Hep
  acc : Accidental := .natural
  deriving BEq, Inhabited
\end{lstlisting}
The pitch $P$ and interval $I$ types obey the following properties, which are
ensured by the \texttt{Scale} type class in Lean:
\begin{enumerate}
\item $I$ is an additive abelian group with some zero interval $0_I \in I$
\item $\forall p \in P, p + 0_I = p$
\item $\forall p \in P, \forall i, j \in I, (p + i) + j = p + (i + j)$
\end{enumerate}
If $p,q \in P$ such that $\exists n \in \mathbb Z. p = q + n \cdot 1_I$, then
$p,q$ are in the same \keyword{tone class}. \texttt{Scale} optionally defines a
list of natural pitches within one $1_I$. There may be tones that do not exist
in any tone class of these natural notes, and they would be considered
accidentals. For example, the natural pitches for a $\pitch{D\sharp}$-minor are
$\pitch{D\sharp}, \pitch{E\sharp}, \pitch{F\sharp}, \pitch{G\sharp}, \pitch{A\sharp}, \pitch{B}, \pitch{C\sharp}$.
The enharmonic $\pitch{E\flat}$-minor have
$\pitch{E\flat},\pitch{F},\pitch{G\flat},\pitch{A\flat},\pitch{B\flat},\pitch{C\flat},\pitch{D\flat}$.
This distinction between enharmonic scales is not representable in 12-tone equal
tempered tuning systems. We can move along the circle of fifth indefinitely,
arriving at exotic keys such as $\pitch{F\sharp^5}$-dorian, which is distinct
from $\pitch{B\flat}$-dorian in some tuning systems.

Under the Pythagorean tuning system \cite{music_a_math}, the tones inside an
interval are established by successively tuning with fifths. For example, the
distance between $\pitch{C}$ and $\pitch{G}$ is tuned to an exact fifth where
the frequency ratio is $2:3$. This process could be repeated to tune all 12
chromatic tones inside an interval. This method is a \emph{lattice of fifths}.
In Pythagorean tuning, the diminished sixth $\pitch{G\sharp}/\pitch{E\flat}$,
which is enharmonic under 12-tone equal temperament as a perfect fifth, is out
of tune as a ``wolf interval''. Prismriver is designed to allow this by not
enforcing an equality between $\pitch{G\sharp}$ and $\pitch{A\flat}$.

The intervals (Listing~\ref{lst:classical-interval-lean}) in heptatonic scale are similarly pairs of $(n, a)$, where $n$ is
a name distance, and $a$ is a semitone distance. A perfect fifth has a name
distance of $4$ and semitone distance of $7$. We can thus distinguish between
augmented second $i_{\pitch{A2}} := (1, 3)$ and minor third
$i_{\pitch{m3}} := (2, 3)$ in this system. We also implemented \keyword{generic
intervals} (called \texttt{KeyInterval}), which has a variable number of
semitones (exactly two by Myhill's property) for each generic interval. For
example, the generic interval with $1$ step $g_1$  in $\pitch{C}$-major maps
$\pitch{C_4}$ to $\pitch{D_4}$ but $\pitch{E_4}$ to $\pitch{F_4}$. Unlike specific intervals, generic intervals are anchored to a certain scale.
\begin{lstlisting}[language=lean,float=htbp,
  caption={Western classical interval examples in Lean \\ (\texttt{Prismriver/Repr/Classical.lean})},
  label={lst:classical-interval-lean}]
-- Example of defining an octave using the Interval type.
def octave : Interval := { name := 7, semitones := 12 }
-- Group actions for the Interval type.
instance : Add Interval where
  add x y := { name := x.name + y.name, semitones := x.semitones + y.semitones }
instance : Sub Interval where
  sub x y := { name := x.name - y.name, semitones := x.semitones - y.semitones }
instance : HMul Int Interval Interval where
  hMul n x := { name := n * x.name, semitones := n * x.semitones }
-- Example of proving additive commutativity given two intervals.
theorem add_comm (x y : Interval) : x + y = y + x := by
  unfold HAdd.hAdd instHAdd Add.add instAdd
  simp
  constructor
  rw [Int.add_comm, Int.add_comm]
\end{lstlisting}
A natural extension of our approach is to use dyadic rationals for accidentals.
This would open the door to quarter tones commonly seen in middle eastern music.
Furthermore, users can contribute their own scale specifiations using
Prismriver's exisiting primitives. The \texttt{Scale} and \texttt{Interval}
interface is sufficiently flexible to represent xenharmonic systems such as
the Bohlen-Pierce scale \cite{music_a_math} where the fundamental interval is an
\emph{tritave} $1:3$ instead of an \emph{octave} $1:2$. Under equal-temperament
tuning, a tritave is divided into 19 tones instead of the 12 in octaves.

\begin{figure}[h]
  \centering
  \includegraphics[width=.4\textwidth]{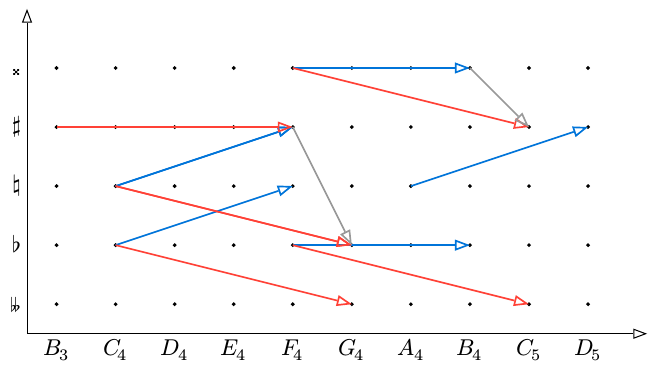}
	\caption{Action of \textcolor{Cerulean}{augmented fourth $i_{\pitch{A4}}$}
and \textcolor{red}{diminished fifth $i_{\pitch{D5}}$} on classical pitches. The
	combination of them produces a \emph{diminished second}, which relates
	enharmonic notes.}
\end{figure}

\begin{figure}[h]
  \centering
  \includegraphics[width=.4\textwidth]{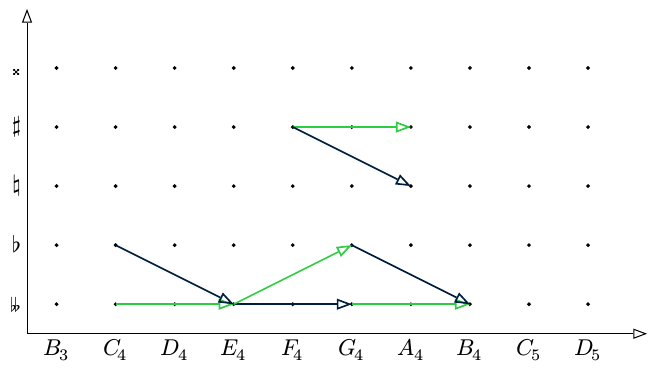}
	\caption{Lattice formed by \textcolor{green}{major third $i_{\pitch{M3}}$} and \textcolor{NavyBlue}{minor third $i_{\pitch{m3}}$}}
\end{figure}

A scale is also required to output a list of pitches in the 0th fundamental
interval, e.g.
$\pitch{D}, \pitch{E}, \pitch{F}, \pitch{G}, \pitch{A}, \pitch{B\flat}, \pitch{C}$.
We can thus recover the key signature from this list.

\subsection{Transpose Action}

In this section we demonstrate usage of Prismriver to model existing music theory and its compatability with Mathlib, furthermore we show that Prismriver's generalized primitives can be used to prove properties beyond an assumed tuning system. A pillar of music theory is the concept of chords and chord progressions,
where music moves from chord to chord. A method to model chord progressions is
via dihedral actions \citep{crans_musical_2008}, where a dihedral action on the circle
of fifths maps one chord to another chord. Pitch classes are represented as elements of $\mathbb{Z}_{12}$, while transpositions and inversions are represented by the dihedral group $D_{12}$ acting on $\mathbb{Z}_{12}$. We define (Listing~\ref{lst:transposeaction-lean}) \keyword{transpose
action}, a generalized version of dihedral actions which can act on arbitrary
scales. A transpose action $a$ is one of
\begin{itemize}
\item $r_i$ where $i \in I$, which offsets a pitch $r_i(p) := p + i$\footnote{In Lean, to prevent operator collision, we use $\cdot$ and $/$ to represent the addition (action) and subtraction of pitches and intervals.}.
\item $sr_{o,i}$ where $o \in P, i \in I$, which maps
	$sr_{o,i}(p) := o - (p - o) - i$. This reflects a pitch about another pitch
	with an offset action.
\end{itemize}
\begin{lstlisting}[language=lean,float=htbp,caption={TransposeAction implementation in Lean. \\ \texttt{(Prismriver/Repr/Dihedral.lean)}},label={lst:transposeaction-lean}]
inductive TransposeAction where
  /-- p → p + i -/
  | r (i : I)
  /-- p → a - (p - a) - I -/
  | sr (a : P) (i : I)

def rInterval (i : I) (p : P) : P := i • p
def srInterval (i : I) (a : P) (p : P) : P := (- (p / a) - i) • a

instance : SMul (@TransposeAction P I) P where
  smul t p := match t with
    | .r i => rInterval i p
    | .sr a i => srInterval i a p
\end{lstlisting}
The function composition of two transpose actions $a_1,a_2$ is a transpose
action. We proved (Listing~\ref{lst:appendix-transposeaction-compose-lean}) the following compositions in Lean:
\begin{equation}
\label{eq:transpose-action-composition}
\begin{aligned}
r_i \circ r_j &= r_{i+j}, \\
r_i \circ sr_{a,j} &= sr_{a,j-i}, \\
sr_{a,i} \circ r_j &= sr_{a,i+j}, \\
sr_{a,i} \circ sr_{b,j} &= r_{2(a/b)+j-i}.
\end{aligned}
\end{equation}

\begin{figure}[h]
  \centering
  \includegraphics[width=.4\textwidth]{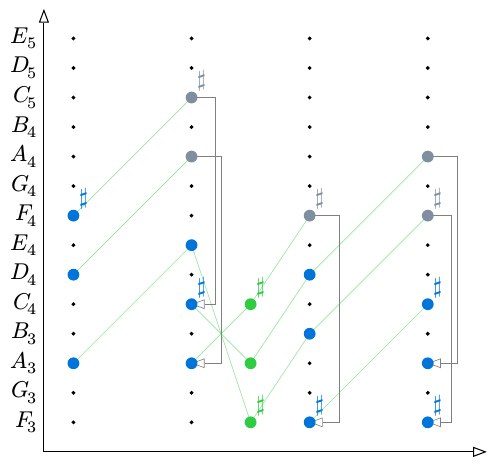}
	\caption{Chord progression of Pachelbel, Canon in D, analyzed in terms of transpose actions}
	\label{fig:Canon in D}
      \end{figure}

In Figure~\ref{fig:Canon in D}, we analyze the
chord progression of Canon in D in terms of transpose actions and chord
inversions.

\cw{After using Prismriver's theory representation library (\texttt{Prismriver/Repr}) to implement and prove the generalized dihedral actions, we model (Listing~\ref{lst:zmod12-dihedral-group-lean}) \citep{crans_musical_2008} by importing \texttt{DihedralGroup} from Mathlib. We define a \texttt{Triad} product type to represent major and minor triads, and use $\mathbb{Z}_{12}$ to represent the equal temperament 12 pitch class. In Prismriver a chord is a list of pitches, so chords of any length are representable. The triad restriction of two parities is due to modeling the dihedral group, not a limitation of Prismriver. Given a pitch class $p \in \mathbb{Z}_{12}$ and interval $i \in \mathbb{Z}_{12}$, a transposition by $i$ semitones is represented by a rotation}
\[
r_i(p) = p + i \pmod{12},
\]

while an inversion is represented by a reflection followed by a rotation,
\[
sr_i(p) = -p + i \pmod{12}.
\]
\begin{lstlisting}[language=lean,float=htbp,
  caption={Implementation of dihedral group action $D_{12}$ on triads in Lean. \\ \texttt{(Prismriver/Theory/Dihedral.lean)}},label={lst:zmod12-dihedral-group-lean}]
inductive Parity : Type | major | minor deriving DecidableEq, Repr, Fintype
-- We use parity to describe whether a chord is major or minor.
-- When we invert a chord, its parity flips.
def Parity.flip : Parity → Parity
  | .major => .minor
  | .minor => .major

abbrev Triad := ZMod 12 × Parity

def transposeTriad (n : ZMod 12) (t : Triad) : Triad := (t.1 + n, t.2)

def invertTriad (n : ZMod 12) (t : Triad) : Triad := (-t.1 + n, t.2.flip)

instance : SMul (DihedralGroup 12) Triad where
  smul g t :=
    match g with
    | DihedralGroup.r n =>
        transposeTriad n t
    | DihedralGroup.sr n =>
        invertTriad (-n) t
\end{lstlisting}

We prove (Listing~\ref{lst:appendix-zmod12-dihedral-group-compose-lean}) the same composition laws from \ref{eq:transpose-action-composition} for $D_{12}$.

We prove in \path{Prismriver/Theory/Dihedral.lean} that under 12-tone equal tempered
tuning system, transpose action is identical to the dihedral group action in \citep{crans_musical_2008}.
Let $\Phi$ map \texttt{TransposeAction} t to $D_{12}$ actions,
$\tau_t : \mathrm{Parity} \to \mathrm{Parity}$ be the parity transformation
induced by $t$, and $\operatorname{Triad}(p,q)$ as the triad
constructed from a root pitch $p$ and parity $q$.

To show that the $D_{12}$ group action on triads is equivalent to the transpose actions on pitches, we prove the following:

\[
\Phi(t)\cdot\operatorname{Triad}(p,q)
=
\operatorname{Triad}\bigl(t\cdot p,\tau_t(q)\bigr).
\]
\cw{To prove this theorem we first specify equal temperament tuning in a local instance of \texttt{Scale Pitch Interval} (Listing~\ref{lst:pitch-interval-zmod12-lean}). We use lemmas (Listing~\ref{lst:appendix-pitch-interval-zmod12-lean}) which show that the tranpose action rotation and reflection actions on pitches are preserved after encoding \texttt{Pitch} and \texttt{Interval} to} $\mathbb{Z}_{12}$.
\begin{lstlisting}[language=lean,float=htbp,caption={Pitch and Interval to $\mathbb{Z}_{12}$ implementation in Lean. \\ \texttt{(Prismriver/Theory/Dihedral.lean)}},label={lst:pitch-interval-zmod12-lean}]
local instance : Scale Pitch Interval := diatonic ⟨.c, .natural⟩ .c

def pitch_toZMod12 (p : Pitch) : ZMod 12 :=
  (p.acc.semitones + nameDistance p.name 0 : Int)

def interval_toZMod12 (i : Interval) : ZMod 12 :=
(i.semitones : ZMod 12)

\end{lstlisting}
We implement $\Phi{(t)}$ as a function, \texttt{TransposeAction.toDihedral},
that maps a \texttt{TransposeAction} \texttt{Pitch} \texttt{Interval} to its
corresponding \texttt{DihedralGroup} action on \texttt{Triad}. We implement
${Triad}(p,q)$, \texttt{pitchTriad}, as a constructor that maps a
(\texttt{Pitch},\texttt{Parity}) pair to a \texttt{Triad} by representing the
pitch in $\mathbb{Z}_{12}$ while keeping its parity
(\texttt{.mapParity}).

\begin{lstlisting}[language=lean,float=htbp,caption={Definitions relating transpose actions to dihedral actions on triads in Lean. \\ \texttt{(Prismriver/Theory/Dihedral.lean)}},label={lst:transposeaction-dihedral-action-lean}]
def TransposeAction.toDihedral :
      @TransposeAction Pitch Interval → DihedralGroup 12
    | .r i =>
        DihedralGroup.r (interval_toZMod12 i)
    | .sr a i =>
        DihedralGroup.sr (interval_toZMod12 i - 2 * pitch_toZMod12 a)

def TransposeAction.mapParity : @TransposeAction Pitch Interval → Parity → Parity
  | .r _, q => q
  | .sr _ _, q => q.flip

def pitchTriad (p : Pitch) (q : Parity) : Triad := (pitch_toZMod12 p, q)
\end{lstlisting}
\cw{Using the definitions (Listing~\ref{lst:transposeaction-dihedral-action-lean}) that bridge transpose action to the dihedral action on triads, we are able to formulate the equivalence theorem in Lean (Listing~\ref{lst:appendix-transposeaction-dihedral-equiv-lean}). We use proof by cases on $t$, which correspond to the rotation and reflection cases. The pitch to $\mathbb{Z}_{12}$ lemmas from earlier are used.}
\subsection{Time and Note}

In the most abstract sense, we model time (Listing~\ref{lst:time-lean}) as a \keyword{time type} $T$, which is
an abelian additive group. $T$ must also be ordered. A time type must also specify
an interval equivalent to 1 bar. A note is a pitch-duration pair $P \times T$,
where $P$ is some pitch type. In most use cases, we use a \keyword{measured
time} type, which is the product type $\mathbb Z \times \mathbb R$. The first
element represents the bar number, and the second represents offset within a
bar.

We opted for this representation for two reasons: (1) It should be easy to shift
music by a fixed amount of bars without calculating the total number of beats
accumulated in those bars. (2) Negative offset within a bar is possible for the
sake of analysis (e.g. pickups).

\begin{lstlisting}[language=lean,float=htbp,caption={Prismriver time definitions in Lean. \\ \texttt{(Prismriver/Repr/Time.lean)}},label={lst:time-lean}]
class Time (T : Type) extends Add T, Sub T, Neg T, SMul Int T, Ord T, Inhabited T where
  zero : T
  /- Maximum time within a bar -/
  bar : T := zero
  default := zero

structure MeasuredTime where
  bars : Int := 0
  offset : Rat := 0
  deriving Ord, BEq
\end{lstlisting}


\section{Input/Output}

Prismriver supports interactive music playing using the \lstinline{#play}
directive. Under the hood, this command elaborates the input expression and
spawns an Alda \cite{alda} process to play the music sequence. Alda is a
programming language for interactive music. This enables rapid prototyping of
music in Lean. We chose to use Alda as the MIDI backend due to its ease of use
compared to alternatives such as Timidity++ \cite{timidity}.

For instance, the code below plays the first five notes of the chorus
of Necrofantasia \cite{th07}:

\begin{lstlisting}
import Prismriver
#play [e4 c'4 b4 d4 e2]
\end{lstlisting}

\subsection{Lilypond-like Music DSL}

On the input side, the user can write music snippets using a LilyPond-like
syntax \cite{lilypond}, which is a purely text-based music interface. The syntax
is constructed as follows.

\begin{bnf}
  p : Tone ::= c // d // e // f // g // a // b
  ;;
  ou : OctaveUp ::= ' // octaveUp ' ,
  ;;
  od : OctaveDown ::= , // octaveDown ,
  ;;
  o : Octave ::= // OctaveUp // OctaveDown
  ;;
  t : Duration ::= numeral
  ;;
  n : Note ::= Tone Octave Time
  ;;
\end{bnf}

We chose LilyPond since it is purely text based and its music typesetting syntax
could be easily implemented in Lean. In LilyPond, music could either be written
in relative (to the previous note) or fixed pitch notation. The pitch of a note
in relative mode is dependent on the previous pitch. Prismriver's DSL uses a
fixed pitch convention, where the pitch of one note is only dependent on its
notation. We made this design decision to reduce volatility.

\subsection{MusicXML}

Prismriver supports outputting classical scores in MusicXML form
\cite{musicxml}. This format of music is widely used by music softwares, and can
be consumed by other music software such as LilyPond or MuseScore.

MusicXML has two formats: Time-wise and Part-wise. This is similar to row and
column major forms of storing matrices. A time-wise MusicXML score is stored
such that one time slice may contain multiple instruments, and a part-wise score
is where each instrument part only contains music pertaining to that instrument.
Tools such as LilyPond can only process part-wise scores. Although part-wise
scores are easy for I/O, they are difficult for music generation and analysis,
where it would be conducive to look at multiple instrument parts on the same
time slice. Therefore, Prismriver stores scores in time-wise format and
transposes them to part-wise format during I/O. This forms the basis of the
\lstinline{Score} structure in Prismriver.

\section{Monadic Algorithmic Compsition}

We created a general monadic interface for algorithmic composition. Since
music composition is a non-causal process, we afford the maximum possible
flexibility to a composition algorithm via the \texttt{CompositionT} monad. The
state of this monad tracks a ``current'' time. It has the functions:
\begin{itemize}
\item \texttt{move}: Move the current time cursor
\item \texttt{addEvent}: Add a new event (used for signaling articulation, etc)
\item \texttt{addNote}: Add a new music note with a certain duration
\end{itemize}

We also have a score analysis monad \texttt{Score.foldlM}, where a folding
function accepts a list of ongoing events, including lingering notes and
instantaneous events (e.g. key change). By combining \texttt{Score.foldlM} and
\texttt{CompositionT}, a composition algorithm (Listing~\ref{lst:classical-monad}) can create accompaniments to
existing scores. By using a monadic structure, we enable reasoning of
composition algorithms using Lean's \texttt{LawfulMonad} class.
\begin{lstlisting}[language=lean,float=htbp,
    caption={Example of using the composition monad. \\ (\texttt{examples/Necrofantasia.lean})},
    label={lst:classical-monad}]
def addPianoNote (note: Classical.Note)
  : Classical.CompositionT Id Unit := addNote note (partId? := .some 0)

def compositionM : Classical.CompositionT Id Unit := do
  addPart 0 { instrument? := .some Instrument.violin }
  let t14 : MeasuredTime := mkRat 1 4
  let t11 : MeasuredTime := mkRat 1 1
  addPianoNote ⟨.new .e 4, t14⟩
  addPianoNote ⟨.new .c 5, t14⟩
  addPianoNote ⟨.new .b 4, t14⟩
  addPianoNote ⟨.new .d 4, t14⟩
  move .bar
  addPianoNote ⟨.new .e 4, t11⟩
\end{lstlisting}

\subsection{Counterpoint}

Counterpoint is a musical technique where there are two or more simultaneous melodic lines \citep{gotham_first_species, hutchinson_first_species}. The starting melodic line is a \keyword{cantus firmus} and its \keyword{counterpoint} consists of simultaneous melodic lines composed to match the cantus firmus \citep{gotham_first_species, hutchinson_first_species}. Species counterpoint is a method for teaching composers how to compose counterpoint, containing up to five variants of composing rule sets called species \citep{hutchinson_first_species}.
\begin{figure}[h]
  \centering
  \includegraphics[width=.4\textwidth]{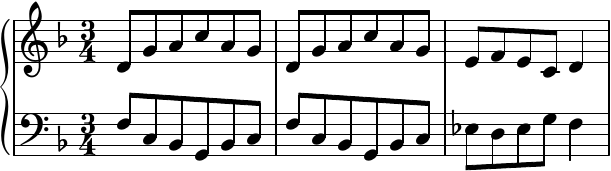}
	\caption{Example of first species counterpoint music notation, generated by Prismriver.}
	\label{fig:counterpoint-notation}
\end{figure}
\subsubsection{First Species Counterpoint}

We implemented an example programmatic interface for a subset of the first species counterpoint composition along with example proofs about its properties to demonstrate using Prismriver for composition. The following subset of First Species rules were implemented using \citep{hutchinson_first_species}:
Let \(c=[c_0,\ldots,c_{n-1}]\) be the list of notes representing the cantus firmus and
  let \(x=[x_0,\ldots,x_{n-1}]\) be the list of pitches representing the counterpoint, with \(n \ge 2\).

  In first species each note of the cantus firmus has a corresponding note from the counterpoint line, thus index \(i\) denotes an aligned pair of the 
  \(i\)-th cantus pitch and the \(i\)-th counterpoint pitch. A \keyword{harmonic interval} is the distance between two notes played at the same time. A \keyword{melodic interval} is the distance between two consecutive notes played one after the other. Let \(H_i=\operatorname{int}(c_i,x_i)\) be the harmonic
  interval at position \(i\), and let
  \(\Delta c_i=\operatorname{int}(c_{i+1},c_i)\) and
  \(\Delta p_i=\operatorname{int}(x_{i+1},x_i)\) be melodic intervals. A \keyword{perfect unison} (\(\mathrm{P1}\)) is when two or more instruments use the same pitch at the same time. A \keyword{perfect octave} (\(\mathrm{P8}\)) is an interval between two notes where one note has twice the frequency of vibration of the other.  Let \(T=\{\mathrm{P1},\mathrm{P8}\}\) be the set of allowed terminal harmonic
  intervals, \(C\) the consonant harmonic intervals, and \(M\)
  the allowed melodic intervals. For a melodic interval \(\Delta v_i\), let \(\operatorname{sgn}(\Delta v_i)\)
  denote its direction, given by the sign of its size in semitones:
  \[
  \operatorname{sgn}(\Delta v_i) =
  \begin{cases}
    +1 & \text{if } \Delta v_i > 0 \quad (\text{the voice ascends}),\\
    \phantom{+}0 & \text{if } \Delta v_i = 0 \quad (\text{the note repeats}),\\
    -1 & \text{if } \Delta v_i < 0 \quad (\text{the voice descends}).
  \end{cases}
  \]
  A \keyword{step} moves to the adjacent note up or down, while a \keyword{leap} skips over one or more notes.
   The allowed melodic intervals \(M\) consist of the steps, the minor
  and major second (\(\mathrm{m2}\), \(\mathrm{M2}\)), along with a restricted
  set of leaps. Writing \(\pm I\) for an interval \(I\) taken in either
  direction and \({+}I\) for its ascending form only,
  \[
  M = \{\pm\mathrm{m2},\, \pm\mathrm{M2},\, \pm\mathrm{m3},\, \pm\mathrm{M3},\,
        \pm\mathrm{P4},\, \pm\mathrm{P5},\, \pm\mathrm{P8}\}
      \;\cup\; \{{+}\mathrm{m6}\},
  \]
  so the allowed leaps are thirds, fourths, fifths, the ascending minor
  sixth, and the octave.
  A \keyword{turning point} is a note where the melody \(v\) changes direction:
  a position \(i\), with \(0<i<n-1\) such that \(\operatorname{sgn}(\Delta v_{i-1})
  \ne \operatorname{sgn}(\Delta v_i)\). A \keyword{tritone} is an interval of 6 semitones. The predicate
  \(\mathsf{NoTritoneTurn}(v)\) holds when no two successive turning points of
  \(v\) are separated by a tritone.
 We model our first-species subset as the predicate
  \(\mathsf{FirstSpecies}(c,x)\), defined by the conjunction of the following
  conditions:
  \[
  \begin{aligned}
  & H_0 \in T,  \\
  & H_{n-1} \in T, \\
  & \operatorname{sgn}(\Delta c_{n-2})
    = -\operatorname{sgn}(\Delta x_{n-2}), \\
  & \forall i < n,\; H_i \in C, \\
  & \forall v \in \{c,x\},\; \forall i < n-1,\;
    \Delta v_i \in M, \\
  & \forall v \in \{c,x\},\; \forall i < n-1,\;
    \Delta v_i \ne 0, \\
  & \forall v \in \{c,x\},\; \mathsf{NoTritoneTurn}(v).
  \end{aligned}
  \]
The rules can be represented as predicates, which can constrain the composition to follow first species. We use the western classical music primitives from \texttt{Repr/Classical} for our counterpoint composition. We define functions (Listing~\ref{lst:counterpoint-comp-lean}) for programatically constructing first species counterpoint, and \texttt{Prop} assertions to compose properties corresponding to the first species rules, so that proving a constructed first species counterpoint is reduced to proving a proposition. Note that since first species counterpoint does not introduce rhythmic variety, we are able to compose counterpoint without representing time. See Listings ~\ref{lst:appendix-counterpoint-comp-lean} and ~\ref{lst:appendix-counterpoint-monad-comp-lean} for more examples of programatically composing counterpoint, and Listing~\ref{lst:appendix-first-species-lean} for more first species proof examples in Lean in Appendix~\ref{appendix-counterpoint}.

\begin{lstlisting}[language=lean,float=htbp,basicstyle=\ttfamily\scriptsize,
  caption={Counterpoint composition examples in Lean. \\ {\footnotesize(\texttt{Prismriver/Composition/Counterpoint.lean})}},
  label={lst:counterpoint-comp-lean}]
def formCounterpointAux : List Pitch → Interval → List Pitch → List Pitch
  | [], _, total => total
  | [x], i, total => (i • x) :: total
  | x :: y :: xs, i, total =>
    let x' := i • x
    let i' := (2 : Int) * (x / y) + i
    formCounterpointAux (y::xs) i' (x'::total)
 -- can programatically define functions to construct counterpoint.
def formCounterpoint (notes : List Pitch) (initial : Interval) : List Pitch :=
  let result := formCounterpointAux notes initial []
  result.reverse
  
#eval formCounterpoint [ (.new .c 4), (.new .d 4), (.new .e 4 .flat), (.new .c 4) ] ((-1) * Interval.p5)
\end{lstlisting}

Since first species counterpoint has multiple rules, each of these rules can be broken down to separate propositions (Listing~\ref{lst:first-species-lean}). This example can be extended to support more species rules by coming up with a generalize interface, where each species ruleset can be an instance of.
\begin{lstlisting}[language=lean,float=htbp,basicstyle=\ttfamily\scriptsize,caption={First species contrary motion example in Lean. \\ {\footnotesize(\texttt{Prismriver/Composition/Counterpoint.lean})}},
  label={lst:first-species-lean}]
-- contraryMotion checks if two intervals move in opposite directions.
def contraryMotion (i j : Interval) : Prop :=
	    (i.semitones > 0 ∧ j.semitones < 0) ∨
    (i.semitones < 0 ∧ j.semitones > 0)
\end{lstlisting}
After defining propositions for each of the rules the propositions can be combined in one \texttt{Prop} function (Listing~\ref{lst:is-first-species-lean}), \texttt{isFirstSpecies}. This way we can abstract away the verification, and one possible extension for this function could be a function parameterized by its species level, and automate which propositions to prove given the species level.
\begin{lstlisting}[language=lean,float=htbp,basicstyle=\ttfamily\scriptsize,caption={Subset of first species ruleset in a Lean proposition example. \\ {\footnotesize(\texttt{Prismriver/Composition/Counterpoint.lean})}},
  label={lst:is-first-species-lean}]
def isFirstSpecies
    (lhs rhs : List Pitch)
    {heq : lhs.length = rhs.length}
    {h : lhs.length ≠ 0}
    (consonant movement : Interval → Prop)
    (beginInterval := allowedTerminalInterval)
    (endInterval := allowedTerminalInterval)
    : Prop :=
    let beginAllowed :=
      let interval := lhs[0] / rhs[0]
      beginInterval interval

    let part1 := ∀ i, ∀ h : i < lhs.length,
      let l := lhs[i]
      let r := rhs[i]
      let interval := l / r
      consonant interval

    let part2 := ∀ i, ∀ h : i < lhs.length - 1,
      let l1 := lhs[i]
      let l2 := lhs[i+1]
      let r1 := rhs[i]
      let r2 := rhs[i+1]
      let lm := l2 / l1
      let rm := r2 / r1
      movement lm ∧ movement rm

    let part3 :=
      allowedLastNotes lhs rhs

    let endAllowed :=
      let interval := lhs[lhs.length-1] / rhs[lhs.length-1]
      endInterval interval

    let part4 :=
      noTritoneTurningPoints lhs ∧ noTritoneTurningPoints rhs

    part1 ∧ part2 ∧ part3 ∧ part4 ∧ beginAllowed ∧ endAllowed
\end{lstlisting}
\subsubsection{Comparison to Other Counterpoint Implementations}
Music Tools \citep{cong_demo_2019} is an Agda library for analyzing and composing music. Originally in Music Tools, rules of first species counterpoint are encoded as type constructors. In a later iteration of Music Tools \citep{leo2022counterpoint}, counterpoint rules are compiled into constraints for an SMT solver, which synthesizes pitches for unfilled positions in a score. In Prismriver, counterpoint is composed explicitly, either as plain lists of \texttt{Pitch} values or through the \texttt{CompositionT} monad. Counterpoint rules are stated as propositions in Lean, and a composition's adherence to counterpoint rules can be proven as a theorem. Mezzo \citep{szamozvancev_well-typed_2017} is a Haskell EDSL that enforces composition rules at compile time. In Mezzo's first species counterpoint implementation, the first species rules are represented as type class constraints. When a rule is violated, compilation fails with a custom type error and its associated violated rule. In Prismriver when a counterpoint rule is violated, the proof on adhering to the violated rule remains incomplete.

\section{Conclusion}

Prismriver is a Lean library for the formalization of music theory and composition of music.
The library contains formalization of chord progressions as dihedral group
action and extends existing music theory formalization works. Prismriver
provides a flexible toolbox for composition and music I/O, compatible with Alda
and MusicXML, where the user is free to define additional tuning and musical
systems. Users can also contribute their own primitives and proofs, leading to
a growing reusable library of music formalization.

Future directions include adding more tuning system primitives such as the
derivation of consonant (e.g. P5) and dissonant (e.g. m1) chords from pitch
overtones, non-western scales, and formalization of chords. Prismriver's score
system is closely modeled after scores in MusicXML and has no innate notion of
stanza and chorus, which could be modeled in a future work. The DSL could be
extended to accommodate additional music notations. On the music analysis side,
a key estimation algorithm could be implemented in a future work using the
existing monadic interface on scores.

\begin{acks}
We thank Shogo Saito (@iehality/Palalansouki) for the Lulu music formalization
library and permission for us to merge the Lulu music widget into Prismriver. We
thank Chris Henson for early discussions on designing the Prismriver
representation library.
\end{acks}
\bibliographystyle{ACM-Reference-Format}
\bibliography{main}
\appendix
\section{Theory Representation}
\subsection{TransposeAction Proofs}
\begin{lstlisting}[language=lean,caption={TransposeAction composition proof examples in Lean. \\ \texttt{(Prismriver/Repr/Dihedral.lean)}},label={lst:appendix-transposeaction-compose-lean}]
theorem rInterval_rInterval (i j : I) (p : P) : rInterval i (rInterval j p) = rInterval (i + j) p := by
  unfold rInterval
  rw [← Scale.add_comm, Scale.smul_assoc]

theorem rInterval_srInterval (i j : I) (a p : P) : rInterval i (srInterval j a p) = srInterval (j - i) a p := by
  unfold rInterval srInterval
  rw [Scale.smul_assoc]
  congr
  have h1 :-(p / a) - (j - i) = -(p / a) + -j + i := by
    have : j - i = j + -i := by rw [Scale.sub_eq_add_neg]
    rw [this, Scale.sub_eq_add_neg, Scale.neg_add, Scale.neg_neg, Scale.add_assoc]
  rw [h1]
  have h2 : -(p / a) - j + i = -(p / a) + -j + i := by rw [Scale.sub_eq_add_neg]
  rw [h2]
\end{lstlisting}

\subsection{Dihedral Group $D_{12}$ Proofs}
\begin{lstlisting}[language=lean,basicstyle=\ttfamily\scriptsize,
  caption={Compositional group action proof examples for dihedral group $D_{12}$ in Lean. \\ \texttt{(Prismriver/Theory/Dihedral.lean)}},label={lst:appendix-zmod12-dihedral-group-compose-lean}]
-- Tm ◦ Tn = Tm+n mod 12, used for r ◦ r.
lemma transposeTriad_add (i j : ZMod 12) (t : Triad) :
    transposeTriad i (transposeTriad j t) = transposeTriad (i + j) t := by
    unfold transposeTriad
    simp
    rw [add_assoc]
    simp
    rw [add_comm]
-- Tm ◦ In = Im+n mod 12, used for r ◦ sr
lemma transposeInverse_add (i j : ZMod 12) (t : Triad) :
  transposeTriad i (invertTriad j t) = invertTriad (i + j) t := by
  unfold invertTriad transposeTriad
  simp
  rw [add_assoc]
  simp
  rw [add_comm]

instance : MulAction (DihedralGroup 12) Triad where
  ...
  mul_smul t x y := by
      cases t <;> cases x
      · -- r.r
        rw [DihedralGroup.r_mul_r]
        simp only [HSMul.hSMul, SMul.smul]
        symm
        apply transposeTriad_add
      · -- r.sr
        rw [DihedralGroup.r_mul_sr]
        simp only [HSMul.hSMul, SMul.smul,sub_eq_add_neg, neg_add_rev, neg_neg]
        symm
        apply transposeInverse_add
  ...
\end{lstlisting}

\subsection{Pitch and Interval to $\mathbb{Z}_{12}$ Proofs}
\begin{lstlisting}[language=lean,caption={Pitch and Interval to $\mathbb{Z}_{12}$ proof examples in Lean. \\ \texttt{(Prismriver/Theory/Dihedral.lean)}},label={lst:appendix-pitch-interval-zmod12-lean}]
lemma pitch_toZMod12_smul_interval (p : Pitch) (i : Interval) :
  pitch_toZMod12 (i • p) = pitch_toZMod12 p + interval_toZMod12 i := by
  unfold pitch_toZMod12 interval_toZMod12
  simp only [smul_Pitch_Interval_acc, smul_Pitch_Interval_name]
  rw [← nameDistance_image (p.name + i.name) p.name 0]
  grind

lemma pitch_toZMod12_smul_neg_interval (p : Pitch) (i : Interval) :
    pitch_toZMod12 ((-i) • p) = pitch_toZMod12 p - interval_toZMod12 i := by
  rw [pitch_toZMod12_smul_interval]
  unfold pitch_toZMod12 interval_toZMod12
  simp only [Interval.instNeg]
  grind
\end{lstlisting}

\subsection{Equivalence of Transpose Actions and Dihedral Actions on Triads Proofs}
\begin{lstlisting}[language=lean,basicstyle=\ttfamily\scriptsize,caption={Equivalence of transpose actions and dihedral actions on triads in Lean. \\ \texttt{(Prismriver/Theory/Dihedral.lean)}},label={lst:appendix-transposeaction-dihedral-equiv-lean}]
theorem transposeAction_toDihedral_triad (t : @TransposeAction Pitch Interval) (p : Pitch) (q : Parity) :
  TransposeAction.toDihedral t • pitchTriad p q = pitchTriad (t • p) (TransposeAction.mapParity t q) := by
  cases t
  apply Prod.ext
  unfold pitchTriad TransposeAction.toDihedral HSMul.hSMul instHSMul SMul.smul
  unfold instSMulDihedralGroupOfNatNatTriad transposeTriad instSMulTransposeAction rInterval
  simp
  rw [pitch_toZMod12_smul_interval]
  unfold pitchTriad TransposeAction.toDihedral TransposeAction.mapParity
  unfold HSMul.hSMul instHSMul SMul.smul instSMulDihedralGroupOfNatNatTriad transposeTriad
  simp
  apply Prod.ext
  unfold pitchTriad TransposeAction.toDihedral HSMul.hSMul instHSMul SMul.smul
  unfold instSMulDihedralGroupOfNatNatTriad invertTriad instSMulTransposeAction
  simp
  rw [add_comm]
  rename_i a
  rename_i i
  have : 2 * pitch_toZMod12 i - interval_toZMod12 a + -pitch_toZMod12 p = 2 * pitch_toZMod12 i - pitch_toZMod12 p - interval_toZMod12 a := by
    unfold pitch_toZMod12 interval_toZMod12
    grind
  rw [this]
  symm
  rw [pitch_toZMod12_srInterval]
  unfold pitchTriad TransposeAction.toDihedral TransposeAction.mapParity HSMul.hSMul instHSMul SMul.smul instSMulDihedralGroupOfNatNatTriad invertTriad
\end{lstlisting}
\section{Counterpoint}
\label{appendix-counterpoint}
\subsection{Counterpoint Composition Lean Code}
In this example, a user can programatically create lists of the \texttt{Pitch} type to compose melodic lines for counterpoint.
\begin{lstlisting}[language=lean,basicstyle=\ttfamily\scriptsize,
  caption={Example of counterpoint composition in Lean using lists of \texttt{Pitch}. \\ {\footnotesize(\texttt{Prismriver/Composition/Counterpoint.lean})}},
  label={lst:appendix-counterpoint-comp-lean}]
def myCantus : List Pitch :=
    [Pitch.new .c 4, Pitch.new .d 4, Pitch.new .e 4, Pitch.new .c 4]
def myCounterpoint : List Pitch :=
    [Pitch.new .c 5, Pitch.new .b 4, Pitch.new .g 4, Pitch.new .c 5]
\end{lstlisting}

A user can also use Prismriver's monadic interface to compose Counterpoint. Below is an example.

\begin{lstlisting}[language=lean,basicstyle=\ttfamily\scriptsize,
  caption={Example of counterpoint composition in Lean using the \texttt{CompositionT} monad. \\ (\texttt{Prismriver/examples/Counterpoint.lean})},
  label={lst:appendix-counterpoint-monad-comp-lean}]
def addPianoNote (note: Classical.Note) (partId : PartId)
  : Classical.CompositionT Id Unit := addNote note (partId? := .some partId)

def cantusM : Classical.CompositionT Id Unit := do
  addPart 0 { instrument? := .some Instrument.acoustic_grand }
  let t14 : MeasuredTime := mkRat 1 4
  addPianoNote ⟨.new .c 4, t14⟩ 0
  addPianoNote ⟨.new .d 4, t14⟩ 0
  addPianoNote ⟨.new .e 4, t14⟩ 0
  addPianoNote ⟨.new .c 4, t14⟩ 0

def counterM : Classical.CompositionT Id Unit := do
  addPart 1 { instrument? := .some Instrument.acoustic_grand }
  let t14 : MeasuredTime := mkRat 1 4
  addPianoNote ⟨.new .c 5, t14⟩ 1
  addPianoNote ⟨.new .b 4, t14⟩ 1
  addPianoNote ⟨.new .g 4, t14⟩ 1
  addPianoNote ⟨.new .c 5, t14⟩ 1
/--
Usage:

`lake env lean --run examples/Counterpoint.lean | alda import -i musicxml | alda play`
-/
def main : IO UInt32 := do
  let t1 := Task.spawn fun _ => compose cantusM |>.run
  let t2 := Task.spawn fun _ => compose counterM |>.run
  let score := (Score.merge (← IO.wait t1) (← IO.wait t2))
  IO.eprintln s!"{score}"
  let xml := score.toMusicXML
  IO.FS.writeFile "counterpoint.xml" (toString xml)
  IO.println s!"{xml}"
  return 0
\end{lstlisting}
\subsection{First Species Counterpoint Proofs}
We provide the rest of the Lean code on proving properties related to a subset of the first species counterpoint that we formalized. These examples follow from Listing~\ref{lst:first-species-lean}.
\begin{lstlisting}[language=lean,basicstyle=\ttfamily\scriptsize,caption={First species counterpoint proof examples in Lean. \\ {\footnotesize(\texttt{Prismriver/Composition/Counterpoint.lean})}},
  label={lst:appendix-first-species-lean}]
-- contraryMotion checks if two intervals move in opposite directions.
def contraryMotion (i j : Interval) : Prop :=
	    (i.semitones > 0 ∧ j.semitones < 0) ∨
    (i.semitones < 0 ∧ j.semitones > 0)
-- allowedLastNotes checks if the final two notes of the melodic lines move in opposite directions.
-- allowedLastNotes can be combined with a proposition that checks if the last note is in perfect octave or unison.    
def allowedLastNotes (lhs rhs : List Pitch) : Prop :=
    lhs.length > 1 ∧ rhs.length > 1 ∧
    let i := lhs.length - 2
    let j := rhs.length - 2
    contraryMotion
      (lhs[i+1]! / lhs[i]!)
      (rhs[j+1]! / rhs[j]!)
 
 example : allowedLastNotes [Pitch.new .d 4, Pitch.new .c 4] [Pitch.new .b 3, Pitch.new .c 4] := by
  unfold allowedLastNotes
  unfold contraryMotion
  simp
  right
  constructor
  native_decide
  native_decide

-- allowedIntervalMovement checks if the given interval is a step or if it's an allowed leap.
def allowedIntervalMovement (i : Interval) : Prop :=
	  isStep i ∨ isAllowedLeap i
 
example : allowedIntervalMovement Interval.mi2 := by
  unfold allowedIntervalMovement
  constructor
  unfold isStep
  simp
\end{lstlisting}
\clearpage

\end{document}